\documentclass[conference]{IEEEtran}
\IEEEoverridecommandlockouts
\usepackage[T1]{fontenc}
\usepackage[dvips]{graphicx}
\usepackage{amssymb}
\usepackage{graphicx}
\usepackage[cmex10]{amsmath}
\usepackage{amsthm}
\usepackage{color}
\usepackage{algorithm}
\usepackage[noend]{algpseudocode}
\usepackage{array}
\usepackage{eqparbox}
\usepackage{diagbox}
\usepackage[tight,footnotesize]{subfigure}
\usepackage{float}
\usepackage{url}
\usepackage{paralist}
\usepackage{multirow}
\usepackage{wrapfig}
\usepackage{color}
\usepackage{tikz}
\usetikzlibrary{trees}
\usetikzlibrary{shapes.geometric, arrows}
\usepackage{cite}
\usepackage{atbegshi}% http://ctan.org/pkg/atbegshi

\def\BibTeX{{\rm B\kern-.05em{\sc i\kern-.025em b}\kern-.08em
    T\kern-.1667em\lower.7ex\hbox{E}\kern-.125emX}}
 
\newtheorem*{remark}{Remark}
\newtheorem{proposition}{Proposition}

\newcommand{\RN}[1]{%
	\textup{\uppercase\expandafter{\romannumeral#1}}
}
\newcommand\thefont{\expandafter\string\the\font}
\allowdisplaybreaks
\begin{document}

\title{Loss Attitude Aware Energy Management for Signal Detection\\
%\thanks{This work was supported by NSF under Grant ENG 1609916.}
}

\author{\IEEEauthorblockN{Baocheng Geng$^a$, Chen Quan$^b$, Tianyun Zhang$^c$, Makan Fardad$^b$ and Pramod K. Varshney$^b$}
\IEEEauthorblockA{$^a$\textit{Department of Computer Science, University of Alabama at Birmingham}\\
$^b$\textit{Department of Electrical Engineering and Computer Science, Syracuse University}\\
$^c$\textit{Department of  Computer Science, Cleveland State University}\\
}
	\thanks{This research was supported by the National Science Foundation under award CAREER CMMI-1750531.}
%\and
}

			%line 2: name of organization, acronyms acceptable\\
			%line 3: City, Country\\
			%:
%			\{bageng, qli33, varshney\}@syr.edu}
	%\author{Nianxia Cao}
	%\thanks{This work was supported in part
% by NSF under Grant ENG 1609916, and in part by AFOSR under Grants FA9550-17-0313 and FA9550-16-1-0077.}
%}
\maketitle

\begin{abstract}
This work considers a Bayesian signal processing problem where increasing the power of the probing signal may cause risks or undesired consequences.
We employ a market based approach to solve energy management problems for signal detection while balancing multiple objectives. In particular, the optimal amount of resource consumption is determined so as to maximize a profit-loss based expected utility function. Next, we study the human behavior of resource consumption while taking individuals' behavioral disparity into account. Unlike rational decision makers who consume the amount of resource to maximize the expected utility function, human decision makers act to maximize their subjective utilities. 
%In resource consumption problems, the  loss attitude characterizing humans' asymmetric valuation towards gains and losses, is the main reason that cause the humans' subjective utilities to be distorted.  
We employ prospect theory  to model humans' loss aversion towards a risky event. The amount of resource consumption that maximizes the humans' subjective utility is derived to characterize the actual behavior of humans. It is shown that loss attitudes may lead the human to behave quite differently from a rational decision maker. %Simulations are provided for illustration and performance comparison.

%Humans invest their resource in performing detection tasks of interest. In a signal detection problem, we establish the relationship between the worker's expected revenue and his/her energy consumption. An optimal amount of energy consumption is derived to maximize the worker's expected utility. Unlike rational decision makers who take actions to maximize their expected utility, humans maximize their perceived utilities with cognitive limitations. Prospect theory is explored to model the workers' loss attitudes and predict their behavior in resource investment. Finally, we define the generalized logarithmic utility function (GLUF) to model the humans' loss attitudes towards being bankrupted, i.e., losing all of his/her energy. Under GLUF, we find that people should invest only a fraction of total resource each time to reduce the risk of severe short term loss.
\end{abstract}


\section{Introduction}

In the areas of wireless communications, target tracking, spectrum sensing, among others, signal detection has been extensively studied in both centralized and distributed settings \cite{varshney2012distributed,viswanathan1997distributed}. Recently, the application of signal detection theory has expanded from traditional hypothesis testing problems to   emerging paradigms such as  crowdsourcing, internet of things (IoT) and human in the loop decision making systems \cite{geng2021utility,geng2018decision,9413745,9747866,sriranga2020human,geng2019decision,9443353}.
%In particular, crowdsourcing enables  a  new  framework to  exploit distributed  human  wisdom  to  solve  problems  that are hard for machines but are easy for humans, such as image labelling and voice transcription. An IoT ecosystem employs web-enabled smart devices that use embedded processors, sensors and communication hardware to collect, send and act on data they acquire from the environments.
The objective of signal detection in most existing literature is to improve the detection accuracy under some resource constraints such as energy\footnote{For the ease of conceptual illustration,  we will use the terms ``resource'' and ``energy'' interchangeably in this paper.} consumption, bandwidth limitation and monetary budget.
For example, power/spectrum allocation problems for centralized, distributed and cluster-based architectures in the context of wireless sensor networks and cognitive radio networks have been studied in \cite{quan2020novel,tounsi2017efficient,shah2014spectrum,quan2021strategic,quan2022enhanced,quan2022efficient,quan2022reputation}. Moreover, budget constrained sensor selection and task scheduling problems have  been studied in the context of crowdsourcing and IoT systems \cite{9049010,cao2019optimal,ibrahim2018data}.
%For example, the authors in \cite{quan2020novel,zhang2018optimal,tounsi2017efficient} have investigated energy/spectrum allocation problems for centralized, distributed and cluster-based architectures in the context of wireless sensor networks and cognitive radio networks. A Budget constrained task scheduling problem was studied in crowdsourcing environments to improve the classification performance \cite{tran2015crowdsourcing,shah2015double}.  
In these above tasks that employ humans/sensor networks to monitor the presence or absence of a phenomenon of interest (PoI), typically all available resources are consumed so that the detection performance is maximized. 
%In practical decision making systems where the participants are selfish humans,scenarios where e, however, 
However, in practical decision making systems such as crowdsourcing and IoT where human participants are selfish and strategic, the costs associated with resource consumption should also be considered. Due to the trade-off between the system performance improvement and cost of the resource, it might not be worthwhile to consume all the available resources.

There are two types of costs associated with energy consumption in the detection and estimation process. One type is the direct energy cost that is needed to generate the probing signal in active sensing systems such as electromagnetic, infrared, pulsed laser and X-ray. Another type of cost, which in general is more significant than the first type, is due to the {\it undesired consequences} caused by the probing signal emitted from ``us''. For instance, in submarine combat scenarios, the active sonar system from us emits a sound wave to detect the range, bearing, and relative motion of the adversary. As the power of the sonar signal increases, although the probing performance improves, at the same time our
states (e.g., location, waveform, beam orientation/aperture) are
more likely to be detected by the adversary, making our system
more vulnerable. Another example is medical imaging such as computed tomography (CT) scans and bone density tests. A large amount of radiation dose results in higher quality images of the human body structure. However, intensive radiation exposure is harmful to the human body. In these environments where the probing signal may cause undesired consequences, the optimal amount of energy consumption must be appropriately determined to balance different objectives.

%In many adversarial or non-cooperative scenarios for detection and estimation, the {\it indirect\footnote{We refer this portion of cost as {\it indirect} in contrast to the {\it direct} cost that is needed to generate the desired signal.}} cost (or risk) is significant due to the fact that the probing signal may cause dangers or reveal the true state of ``us'' to the ``adversary''. For instance, in air or undersea combat scenarios, we may use a drone/UAV or an electromagnetic signal to probe an adversary's multi-function radar system. As we increase the power of the emitting signal, although the probing performance improves, our states (e.g., location, waveform, beam orientation/aperture) are more likely to be detected by the adversary, making our system more vulnerable. Another example is medical imaging such as computed tomography (CT) scans and bone density tests. A large amount of radiation dose results in higher quality images of the human body structure. However, intensive radiation exposure is harmful to the human body. In these adversarial environments, the optimal amount of energy consumption must be appropriately determined to balance different objectives.

This paper employs a utility based approach where we consider the profit to be the detection performance and the loss to be the energy consumption and undesired consequences caused by the probing signal,
%\footnote{The terms `resource' and `energy' will be used interchangeably.}
i.e., we consider a profit and loss proposition. The objective is to determine the optimal amount of energy consumption to balance the trade-off between system performance gain and energy consumption cost. In most signal detection problems, the detection accuracy improves as more energy is consumed. However, in our setting, the rate of detection accuracy increment slows down as more energy is spent, i.e, the phenomenon of `diminishing returns' sets in. After a saturation point, the extra profit obtained by spending  additional units of energy is lower than the cost incurred. In this case, the usage of additional amount of energy is neither useful nor advisable.

%the `return on investment' is lower than the cost per unit resource consumption, and it makes the usage of additional resources not legitimate.

%In the first part of this work, we study how the rate of detection accuracy increment behaves as more energy is consumed and determine the optimal amount of energy usage to maximize the system performance based on expected utility theory (EUT). Under the framework of EUT, it is assumed that decisions are made  on an expected value or linear utility basis \cite{mongin1997expected}. However, according to psychology studies, the determination of the value of something must not be based on its explicit price/profit, but rather on the implicit utility it yields. While the price/profit is equal for everyone; the utility, however, is dependent on the particular circumstances of the human making the estimate \cite{kahneman2013prospect}.

Moreover, we consider that the decision on energy consumption is made by a human decision maker and we aim to analyze the behavioral difference on energy consumption across individuals. Unlike rational decision makers who perceive the expected utilities precisely, humans' perception of utilities of an event is subjective and distorted due to cognitive biases. According to psychology studies, one prominent feature of human cognitive biases is their loss attitude that characterizes the asymmetric valuation towards gains and losses \cite{kahneman2013prospect}.
%aversion as one strongly prefers avoiding losses than achieving gains. 
In order to incorporate human loss attitude in the analysis, we adopt  prospect theory (PT) to carry out our analysis. The Nobel-prize-winning prospect  theory  provides  an  accurate  description  of loss aversion in human decision making by utilizing a value function to describe humans' asymmetric valuation towards gains and losses. There have been several works that incorporated PT into hypothesis testing to model human decision making \cite{geng2020prospect,geng2019amelioration,9133140,geng2021collaborative,geng2021augmented}. 
%PT has been exploited to model human behavior in \cite{nadendla2016towards}, the authors applied PT to hypothesis testing and analyzed the behavior of two types of players: optimists and pessimists.
The authors in \cite{yu2014spectrum,yousefvand2018impact,li2014users} employed PT to model human choice for goods and service exchange in the applications of spectrum sensing as well as wireless sensor networks.

%has also been used in spectrum sensing \cite{yu2014spectrum}, as well as studying the user's behavior in wireless networks \cite{yousefvand2018impact,li2014users}. 

To the best of our knowledge, no previous work has investigated the loss aversion aware energy consumption problem in the context of signal detection in general and for hypothesis testing problems in particular. In this paper, we first formulate a profit-loss based expected utility function in signal detection and determine
the optimal amount of resource consumption while balancing multiple objectives. 
%In this paper, we first explore how the system performance improves as more energy is consumed and determine the optimal amount of energy consumption to balance multiple objectives in the expected utility framework. 
Next, we employ the value function from PT  to construct the subjective utility functions and  characterize the humans' loss aversion behavior of energy consumption. We study the optimal strategy of energy usage that maximizes human subjective utility under PT with fixed and weighted average reference points, respectively. %Moreover,  a generalized log utility function is developed to capture humans' different loss attitudes towards the cost of energy consumption. 

\section{Problem Formulation}

Consider a binary signal detection problem and our goal is to decide on the amount of energy consumption while solving the problem.  Let the amount of energy consumed to perform the detection task be denoted by $p$.  The probability of making a correct decision is a function of $p$ and is denoted by $D(p)$. In a number of applications, $D(p)$ is increasing and concave with respect to $p$, i.e, the rate of increase of $D(p)$ slows down as $p$ increases. 

%This phenomenon is analogous to the law of diminishing returns in economics, which states that at some point, the increase of output declines when adding one or more factors of production \cite{marshall2009principles}. 

\subsection{Examples that illustrate the concavity of $D(p)$}

In this subsection, we provide two examples to show that $D(p)$ satisfies the above mentioned property in a wide range of signal detection problems.

{ \textit{1. Binary hypothesis testing with shift of means.}} The shift-of-mean hypothesis testing problem characterizes a large number of problems in signal processing and communications. Under the two hypotheses, the observation is assumed to be a Gaussian random variable with means  $\pm \sqrt{p}$ and the same variance $\sigma^2$. Consider that in order to generate the signals with amplitude $\pm \sqrt{p}$, the required power is $p$. The optimal Bayesian detector \cite{varshney2012distributed} has the  probability of successfully detecting the hypothesis given by 
%uncoded binary phase shift keying BPSK symbols are transmitted over communication channels corrupted by Gaussian noise. If we employ the transmission power $p$, the probability of successfully detecting the signal, i.e., correctly predicting which signal is transmitted, is 
\begin{equation}\label{eq:dp}
 D(p) = 1 - Q(\sqrt{
    \frac{p}{\sigma^2}})   
\end{equation}where $Q(t)$ is the probability that a standard normal random variable takes a value larger than  $t$: $Q(t) = \frac{1}{\sqrt{2\pi}}\int_{t}^{\infty}\exp (-\frac{u^2}{2})du$.

%2. \textit{Asymptotic hypothesis testing.} In binary hypothesis testing problems, there are two hypotheses denoted by $H_0$ and $H_1$ where the observation $X$ follows probability density functions (pdfs) $P_0$ and $P_1$, respectively. Based on an $n$ observation sequence $X^n$ drawn i.i.d from one of the two  distributions, the Bayesian probability of correctly detecting the hypothesis is given by $D(n) \approx 1- 2^{-n c^*(P_0,P_1)}$, where $c^*(P_0,P_1)=-\log \min_{\lambda\in(0,1)}\int P_0^{\lambda}(x)P_1^{1-\lambda}(x)dx$, which is also known as the Chernoff information, is the best achievable exponent for Bayesian probability of error \cite{cover2012elements}.

{\textit{2. Asymptotic binary hypothesis testing.}} In binary hypothesis testing problems, there are two hypotheses where the observation $X$ follows a probability density functions (pdf) $P_0$ and $P_1$ under each hypothesis. Based on an $n$ observation sequence $X^n$ drawn i.i.d from one of the two  distributions, the Bayesian probability of correctly detecting the hypothesis is given by $D(n) \approx 1- 2^{-n c^*(P_0,P_1)}$, where $c^*(P_0,P_1)=-\log \min_{\lambda\in(0,1)}\int P_0^{\lambda}(x)P_1^{1-\lambda}(x)dx$, known as the Chernoff information, is the best achievable exponent for Bayesian probability of error \cite{cover1999elements}. Let the number of observations $n$ denote the amount of energy consumption, $D(n)$ is an increasing and concave function with respect to $n$.

%2. \textit{Collaborative human decision making.} There are certain types of tasks, e.g., image annotation and text transcription, that are easy to solve by humans but hard for machines. Consider a collaborative human decision making structure that consists of a group of $n$ human experts, where each human has an expected probability $p_c>0.5$ of making correct decisions in binary decision making. A project manager collects the local decisions made by the $n$ humans and makes the final decision by employing the majority rule, i.e., selecting the choice that has more than half of the votes. It was shown in \cite{geng2018decision} that the probability that the project manager makes the correct decision is increasing and concave with respect to the group size $n$. 

%4. \textit{Applications in deep learning.} Recently, deep neural networks (DNN), convolutional neural networks (CNN). etc.,  have become powerful tools for classification by using multiple layers to progressively extract high level features from raw inputs \cite{zhang2019fusion}.  If we treat the computational efforts that include the number of input raw data samples, the number of layers, the size of each layer and the number of training rounds as the resource consumption, the classification accuracy of these neural network based decision making systems is expected to be an increasing and concave function as more computational resources are used.

\subsection{Expected utility as a function of the energy consumption}

In the setting of binary decision making, we consider that the human derives a profit $s$ if he/she makes the correct decision. When the human makes a wrong decision, the profit is set equal to  $0$ after normalization (One may also formulate the problem such that the human has profit $0$ for correct decisions and cost $s$ for wrong decisions.). We use $c$ to denote the cost per unit energy consumption. Note that here $c$ represents the total cost of energy consumption that may consist of  two parts: one is the direct energy cost to generate the signal, and the other part is the undesired consequences caused by $p$ (such as the revelation of states in radar countermeasures and the risks of radiation in medical imaging).

Since the probability of making a correct decision is $D(p)$, the expected utility (EU) function when spending energy $p$ to perform the task is  
\begin{equation}\label{utilityfunc_eu}
U(p)=   s D(p)-cp.
\end{equation}Under the assumption that $D(p)$ is increasing and concave with respect to $p$, and the cost function is linear\footnote{Here, we consider that $p$ is the amount of energy and $c$ is the cost per unit energy, so that the cost function $cp$ is linear in $p$. In applications where $p$ represents  other types of resource consumption, the cost function might be nonlinear, e.g., convex\cite{varian1992microeconomic}.} with respect to $p$, the optimal amount of energy $p^*$ that maximizes $U(p)$ is given by
\begin{equation}\label{pstar}
p^*=(D')^{-1}(\frac{c}{s})\triangleq D^*(\frac{c}{s})
\end{equation}where $D'(\cdot)$ is the first order derivative of $D(p)$ with respect to $p$ and $D^*(\cdot)$ is the inverse function of $D'(\cdot)$. Note that since $D(p)$ is increasing and concave, $D'(\cdot)$ is strictly positive and is a decreasing function with respect to $p$. Hence, its inverse function $D^*(\cdot)$ is a decreasing function as well. As a result, it is readily seen from (\ref{pstar}) that the optimal energy consumption $p^*$ increases as $s$ becomes larger or $c$ becomes smaller.

%In this section, we employed expected utility theory (EUT) and derived the optimal energy usage for typical signal detection problems. Our model can be generalized to a diverse range of classification and detection problems in signal processing, wireless communications, supervised machine learning and economics, where the system utility function is increasing and concave with respect to the amount of energy consumption. Moreover, our method can be used to model individual and social activities in daily life. For example,  when limited time is allocated to achieve multiple objectives, the human may need to determine how much time to spend for a particular task, e.g., preparing for an exam,  when the probability of passing the exam is  increasing and exhibits concavity as more time is consumed for preparation. The nature of $D(p)$ in our model is analogous to the law of diminishing returns in economics, which states that at some point, the increase of the utility functions declines when adding one or more factors of production \cite{marshall2009principles}. 

\section{Loss attitudes modeled by prospect theory}

In the previous section, we analyzed the optimal energy consumption when the human is a rational decision maker, i.e., the human attemps to maximize the expected utility. In practice, however, humans take actions to maximize their subjective utilities, which are distorted due to cognitive biases\cite{kahneman2013prospect}. 
%In signal detection problems, recall that the human perceives the profit $s$ of successful detection of the signal to be gains and energy consumption to be losses. 
In this section, we employ prospect theory to model the human's asymmetric valuation towards gains and losses and study how it affects the energy investment strategy.

Prospect theory (PT), proposed by Kahneman and Tversky in 1979 \cite{kahneman2013prospect}, suggests that people are usually loss averse in the sense that loss feels worse than the gain of an equivalent amount feels good. The value function 
%plotted in Figure 1 
characterizes the phenomenon of loss aversion by assigning a subjective utility $V(x)$ to an outcome $x$:
\begin{align}\label{eq:valuefunc}
V(x)=
\begin{cases}
(x-r)^\lambda, \ {x \geq r}\\
-\beta(r-x)^\lambda, \ {x < r}
\end{cases}
\end{align}where $x$ is the actual gain (when it is positive) or loss (when it is negative). One prominent feature of PT is that the human evaluates outcomes relative to a reference point, and then classifies them as gains and losses. In \eqref{eq:valuefunc}, the reference point is represented using $r$, which is a subjective point that varies from one individual to another. $\beta$ is the loss aversion coefficient, and $V(x)$ reflects people's different loss aversion attitudes  by the variation of parameter $\beta$. When the human is more loss averse, $\beta$ increases and the subjective utility of a fixed value of loss appears to be more significant. $\lambda$ characterizes the phenomenon of diminishing marginal utility, which says that as the total number of units of gain (or loss) increases, the utility of an additional unit of gain (or loss) to a person decreases. %Diminishing marginal utility is reflected in the graph as the curve saturates in both directions. 
According to experimental data collected by Kahneman and Tversky \cite{tversky1992advances}, both $\beta$ and $\lambda$ are positive values and the mean values of $\beta, \lambda$ in the group of subjects are  $2.25, 0.88$, respectively.
%\begin{figure}[htb]	\centering\includegraphics[width=0.6\columnwidth]{newvalue_1.eps}
%	\caption{Value function from prospect theory}
%\end{figure}

In the following, we apply the value function to both gain and loss, and derive the optimal energy consumption strategy that maximizes the human's subjective utility. Since the choice of the reference point determines the form of the value function, and hence, the human's subjective utility, we proceed with the analysis of two different types of reference point models. 
\subsection{Fixed reference point}
First, we consider that the reference point is fixed at $r=0$ so that the value function is given in \eqref{eq:valuefunc}. The profit $s$ if the human correctly detects the signal is perceived to be $V(s)= s^\lambda$, and the loss of consuming energy amount $p$ is perceived to be $V(-cp)= -\beta(cp)^\lambda$. Hence, when the human consumes energy $p$, the subjective utility under PT with fixed reference point 0 is given by:
\begin{align}\label{eq:PTutility_fix}
    U_f(p)=s^\lambda D({p}) -\beta(cp)^\lambda
\end{align}Note that when $\beta=1,\lambda=1$, \eqref{eq:PTutility_fix} reduces to the expected utility theory based maximization problem \eqref{utilityfunc_eu}. The optimal energy that maximizes the humans' perceived utility is given by $p_f=\arg\!\max\limits_{p} \ U_f(p)$.
%\begin{align}
%   p_f^{PT}=\arg\!\max\limits_{p} \ U_{PT}^f(p) 
%\end{align}
Generally, $U_f(p)$ is not necessarily convex. In the derivation of $p_f$ via the above equation, there might exist several local maximum points and we must make a comparison to determine the global optimum.

In the following, we employ the example of shift of mean hypothesis testing where $D(p)$ is given by \eqref{eq:dp} and explicitly derive the expression for $p_f$. In such a case, we show that the objective function \eqref{eq:PTutility_fix} is quasi-concave so that the local maximum point is globally optimal.

\begin{proposition}
In the task of shift-of-mean hypothesis testing, there exists a unique  $p_f$ that maximizes the human's subjective utility under PT with fixed reference point $r=0$. If the value of $\lambda\geq0.5$, $p_f$ decreases as $\beta$ becomes larger. 
\end{proposition}
\begin{proof}
The first order derivative (FOD) of $D(p)= 1-Q(\sqrt{p/\sigma^2})$ with respect to $p$ is given by $D'(p)=\frac{1 }{2\sigma\sqrt{2\pi {p}}}e^{-\frac{{p}}{2\sigma^2}}$. Hence, the FOD of 
%the objective function of 
\eqref{eq:PTutility_fix} is given by
\begin{align}\label{eq:foc_1}
    \frac{\partial U_f}{\partial p}
    %&=s^{\lambda}D'(p) - \beta c^{\lambda}p^{\lambda -1}\label{eq:foc_1}\nonumber\\&
    =s^{\lambda} \frac{1 }{2\sigma\sqrt{2\pi {p}}}e^{-\frac{{p}}{2\sigma^2}}-\lambda \beta c^\lambda p^{\lambda-1}
\end{align}We solve for the local optimum point $p_f$ by setting the FOD of $U_f$  equal to 0, and get $e ^{-\frac{p_f}{2\sigma^2}}=\frac{2\sqrt{2\pi}\sigma \lambda \beta c^{\lambda}}{s^{\lambda}{p_f}^{0.5-\lambda}}$, which, after some mathematical manipulations, can be expressed as 
%\begin{align}\label{trans1}
%    e ^{-\frac{p_f}{2\sigma^2}}=\frac{2\sqrt{\pi}\sigma \lambda \beta c^{\lambda}}{s^{\lambda}{p_f}^{0.5-\lambda}}
%\end{align}After mathematical transformations, the above equation can be written in the form of 
%Taking the log of both sides in the above equation, we have
%\begin{align}\label{trans2}
%    -\frac{p_f}{2\sigma^2}= \log \frac{2\sqrt{2\pi}\sigma \lambda\beta c^{\lambda}}{s^{\lambda}{p_f}^{0.5-\lambda}}
%\end{align}Moving the term ${p_f}^{0.5-\lambda}$ in \eqref{trans2} from the right hand side to the left hand side gives
%\begin{align}\label{trans3}
%    (0.5-\lambda)\log p_f +\left(\frac{-p_f}{2\sigma^2}\right)=\log \frac{2\sqrt{2\pi}\sigma \lambda\beta c^{\lambda}}{s^{\lambda}}
%\end{align}Dividing $0.5-\lambda$ from both sides of the above equation, we have 
%\begin{align}\label{trans4}
%    \log p_f + \frac{-p_f}{2\sigma^2(0.5-\lambda)}= \frac{\log \frac{2\sqrt{2\pi}\sigma \lambda\beta c^{\lambda}}{s^{\lambda}}}{0.5-\lambda}
%\end{align}Finally, we subtract the term $\log \frac{1}{2\sigma^2 (0.5-\lambda)}$ to both sides of \eqref{trans4} and have
\begin{align}
    \log \left(p_f\frac{1}{2\sigma^2 (\lambda-0.5)}\right)+p_f\frac{1}{2\sigma^2(\lambda-0.5)}= z
\end{align}where $z=(\log \frac{2\sqrt{2\pi}\sigma \lambda\beta c^{\lambda}}{s^{\lambda}})/(0.5-\lambda)+\log (2\sigma^2 (0.5-\lambda))$. Hence, the solution $p_f$ is given by
\begin{align}
    p_f= 2\sigma^2 (\lambda-0.5)\omega(z)
\end{align}where $\omega(\cdot)$ is the Wright Omega function that satisfies the relationship $\log(\omega(x))+\omega(x)=x$. Since $\omega(\cdot)$ is one to one   and strictly increasing function \cite{weisstein2002lambert},  it guarantees that we have a unique solution of $p_f$. Due to the monotonicity of $\omega(\cdot)$, one can show that when $p<p_f$, $\frac{\partial U_f}{\partial p}>0$ and when $p>p_f$, $\frac{\partial U_f}{\partial p}<0$, indicating that $U_f(p)$ is a quasi-concave function and the local maximum point $p_f$ is globally optimal. Note that in the above derivation, we use $\lambda>0.5$. It has been reported in the experiments \cite{tversky1992advances} that the typical value of $\lambda$ that characterize the effect of diminishing marginal utility is $0.88$ and the condition $\lambda> 0.5$ holds.

Next, we note that setting the FOD of the objective function given in \eqref{eq:foc_1} equal to 0 is equivalent to:
\begin{equation}\label{eq:foc_transform}
D'(p){p}^{1-\lambda}=\frac{\beta \lambda c^\lambda }{s^\lambda}
\end{equation}

It is easy to see that $D'(p)p^{1-\lambda}$ is a decreasing function with respect to $p$ when $\lambda> 0.5$. Since the right hand side of \eqref{eq:foc_transform} becomes larger as $\beta$ increases, it is clear that $p_f$ decreases as $\beta$ takes a larger value.
\end{proof}

\subsection{Weighted average reference point}

% When the worker consumes power amount $p$, the stochastic reference point can be written as the weighted average of the maximum possible profit $U$ and the minimum possible profit $-cp$ \cite{long2014prospect,uppari2018modeling}

The reference point in the model of  \cite{long2014prospect,uppari2018modeling} can be  written as the weighted average of the maximum and minimum profits associated with a particular action. In our case where the human consumes $p$ amount of energy to perform the task, the profit of successfully detecting the signal is $s$, and the cost is the energy consumption $-cp$. Thus, the weighted average reference point can be expressed as:
\begin{equation}\label{stochastic_reference}
r(s,p) = t s + (1-t)(-cp)
\end{equation}where $t\in[0,1]$ characterizes the human's level of optimism. When $t$ is large, the human is optimistic and has a high expectation of the signal being correctly detected, while a low value of $t$ suggests that the human is pessimistic and is more likely to expect the detection result to be wrong. If we replace the  reference point 0 by the weighted average reference $r$ in the value function \eqref{eq:valuefunc}, the human's subjective utility of spending energy $p$ is given by:
%With the weighted average reference point given in \eqref{stochastic_reference}, the value function of PT becomes
%\begin{align}\label{eq:valuefunc_s}
%V_s(x)=
%\begin{cases}
%(x-r)^\lambda, \ {x \geq r}\\
%-\beta(r-x)^\lambda, \ {x < r}
%\end{cases}
%\end{align}The human's optimal amount of energy consumption under PT with weighted average reference point is obtained from
\begin{align}
U_w(p)&= (s-r)^\lambda D({p}) -\beta\left(r-(-cp)\right)^\lambda\nonumber\\
&=(s+cp)^\lambda ((1-t)^\lambda D(p) - \beta t^\lambda )\label{eq:utilityPT_stochastic}
\end{align}We assume that the human chooses to spend a certain amount of energy to perform the detection task only when the subjective utility is positive. If the subjective utility is non-positive, the human would rather not participate in the detection task, i.e., chooses to spend 0 energy.

\begin{proposition}
To maximize the subjective utility under PT with weighted average reference points, the human spends all the available energy if $(1-t)^\lambda D(p_0) - \beta t^\lambda > 0$, where $p_0$ is the amount of available energy, and spends 0 energy otherwise.
\end{proposition}
\begin{proof}
Since $u_1(p)=(s+cp)^\lambda$ is a positive term, it is clear that  \eqref{eq:utilityPT_stochastic} is negative when $u_2(p)=(1-t)^\lambda D(p) - \beta t^\lambda \leq 0$. In such a case, the subjective utility function is negative and the human does not participate in performing the detection task. On the other hand when $u_2(p)>0$, the objective function \eqref{eq:utilityPT_stochastic} is positive and increasing with respect to $p$ as both $u_1(p)$ and $u_2(p)$ are increasing functions of $p$. In this situation, the human should consume all of the available energy to maximize the subjective utility. \end{proof}Note that given the parameters $t,\lambda$ and $p_0$, there is a threshold $\beta^s =\frac{(1-t)^\lambda D(p_0)}{t^\lambda}$ so that a human with loss aversion parameter $\beta\leq \beta^s$ chooses to spend everything and a human with $\beta >\beta^s$ chooses to spend nothing.

\begin{remark}
The condition that the human spends 0 energy is equivalent to $(\frac{t}{1-t})^\lambda \beta > D(p_0)$. In other words, given $p_0$ and $\lambda$, the human is more likely to spend 0 energy instead of spending all the energy if the human's optimism towards the detection result $t$ is high and if the loss aversion parameter $\beta$ is large. One may interpret $t$ as another type of loss aversion in the sense that a human is more loss averse if he/she has higher expectation of the detection result to be successful. 
\end{remark}

% Some people may argue that when the worker consumes power $p$ and makes the correct prediction, the actual income is $U-cp$. Thus, if we define the stochastic reference point to be 
% $r'(U,p)= \alpha (U-cp)+(1-\alpha)(-cp)=\alpha U-cp$. Substitute this reference point into the worker's optimization problem (5), we have:
% \begin{small}
%  \begin{align*}
%  \max\limits_{p} \quad PT'_s(p)&= (U-r')^\lambda D(\textbf{p}) -\beta\left(r'-(-cp)\right)^\lambda\nonumber\\
%  &=\left((1-\alpha)U+cp\right)^\lambda
% D(p)-\beta (\alpha U)^\lambda \end{align*}
%  \end{small}
% which is again an increasing function with respect to $p$. In this case, we have a similar result with Lemma 4 that when $PT'_s(p_0)<0$, the workers should not undertake the task and he/she should spend all of the resource otherwise.

\section{Numerical results}

For illustration, we conduct experiments for the scenario where a human spends some amount of energy to perform shift-of-mean hypothesis testing as described in Section II.A. We assume the channel noise variance  to be $\sigma^2=1$. The profit of successfully detecting the signal is $s=40$ and the cost of unit energy consumption is $c=5$. First, prospect theory is employed to model human's loss attitudes where we set  the diminishing marginal utility parameter $\lambda=0.88$ and vary the human's loss aversion parameter $\beta$. In Figure 1 (a), we plot the optimal amount of energy consumption as  $\beta$ changes under EU, PT with fixed reference point and PT with weighted average reference point, respectively. As the probability of correctly detecting the signal by spending energy $p$ is $D(p)$ given in (1), the optimal amount of energy consumption under EU can be solved via (4) and we obtain that $p^*=0.98$. It can be observed  that  under EU, the optimal amount of energy usage is a constant, without being affected by the variation of $\beta$. When PT is incorporated to model human's loss attitudes, the amount of energy consumption is quite different from the results obtained by assuming that humans are rational. Under PT with a fixed reference point $r=0$, the optimal amount of energy consumption $p_f$ monotonously decreases as $\beta$ becomes larger, which corroborates our analysis in Proposition 1. Under PT with a weighted average reference point where we set the optimism parameter to be $t=0.3$, there is a threshold denoted by $\beta^S$ such that if $\beta<\beta^S$, the human spends all the available energy and as the human is more loss averse in the sense that $\beta\geq\beta^S$, the human spends 0 energy. 
\begin{figure}[htb]	\centering\includegraphics[width=1\columnwidth]{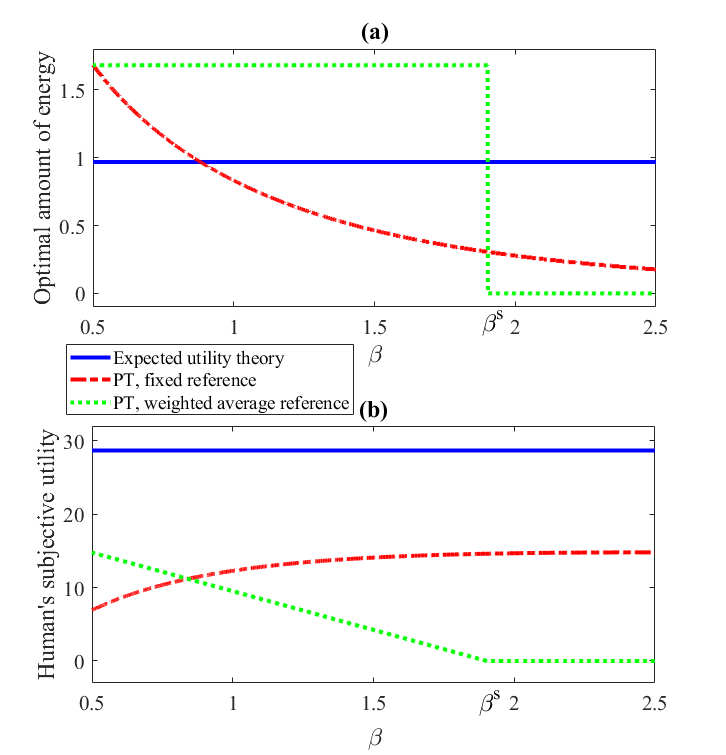}
	\caption{Human behavior of energy consumption under PT}
\end{figure}

In Figure 1(b), we let $\beta$ vary and plot the human's subjective utility of spending the optimal amounts of energy (the optimal amount of energy consumption has been obtained in Figure 1 (a)) when his/her loss attitude is modeled by EU, PT with a  fixed reference point and PT with a weighted average reference point, respectively. %Similarly,  under EUT the subjective utility is a constant when $\beta$ changes. 
It is  observed that the subjective utilities under PT (fixed or weighted average reference point) are smaller than the expected utility. This is because the human perceives the gain to be smaller due to diminishing marginal utility parameter $\lambda$ and perceives the cost to be larger when the  loss aversion parameter $\beta>1$. It is interesting to observe that the human's subjective utility of spending $p_f$ is increasing with respect to $\beta$ under PT with a fixed reference point. The reason is that as $\beta$ increases, $p_f$ becomes smaller. In the subjective utility  function \eqref{eq:PTutility_fix},  though the term $s^\lambda D(p)$ representing the gain becomes smaller, the perceived cost $\beta (cp)^\lambda$ decreases as well. It is the domination of the second term over the first one that leads to the increase of subjective utility as $\beta$ becomes larger. On the other hand, under PT with weighted average reference points, the optimal amount of energy consumption $p_s$ does not change when $\beta$ increases except when the threshold $\beta^S$ is reached, and hence, the subjective utility is decreasing as the human is more loss averse. Note that there is an abrupt change when $\beta$ reaches $\beta^S$ where the human changes the energy consumption strategy from spending all the available energy  to spending nothing.

\section{Conclusion}
In this paper, we investigated the use of utility theory to optimize resource consumption for signal detection problems when the probing signal of the active sensing system may cause undesired consequences. Under the expected utility framework, the optimal amount of resource consumption was derived that maximized a profit-cost based objective function. 
We further considered that the decision makers are humans and study how loss attitude impacts the actual human behavior of resource consumption. Prospect theory was employed to model humans' asymmetric valuation towards gains and losses. The characterization of human behavioral properties in resource consumption is not only important to analyze the human decision quality in non-cooperative classification/detection tasks but also relevant in areas like behavioral informatics, task allocation and  incentivization in crowdsourcing and IoT systems. In the future, it will be worthwhile to analyze the impacts of more complicated cost functions, e.g., non-linear, on human decision making strategies in resource consumption.

\bibliography{refer}
\bibliographystyle{IEEEbib}

% 	\end{proof}
% \end{appendices}

\end{document}